\documentclass[letterpaper, 10 pt, conference]{ieeeconf}  
\usepackage{amsmath,amsfonts,bm}
\usepackage{algorithmic}
\usepackage{algorithm}
\usepackage{array}
\usepackage{CJKutf8}                    
\usepackage{graphicx}
\usepackage[dvipsnames]{xcolor}      
\usepackage{booktabs}
\usepackage{array}
\usepackage{verbatim}
\usepackage{cite}
\usepackage{url}
\IEEEoverridecommandlockouts                              
\title{\LARGE \bf
Deep Photonic Reservoir Computer Meets UAV Control: An ultra-fast learning-based compensator for agile flight in confined space}

\author{Qinxiao Ma\textsuperscript{1*}, Ruiqian Li\textsuperscript{1*}, Cheng Wang\textsuperscript{2\dag}, and Yang Wang\textsuperscript{1\dag}\thanks{*Equal contributions; \dag~Co-corresponding authors. This work was supported by the National Natural Science Foundation of China (Grant Nos. 62503329) and by the Science and Technology Commission of Shanghai Municipality (Grant Nos. 24JD1402400 and 24TS1401500).}
}
\begin{document}

\maketitle
\footnotetext[1]{Q. Ma, R. Li and Y. Wang are with the School of Information Science and Technology, ShanghaiTech University, Shanghai 201210, China (e-mail: \{maqx2023, lirq2022, wangyang4\}@shanghaitech.edu.cn).}
\footnotetext[2]{C. Wang is with the School of Information Science and Technology and the Shanghai Engineering Research Center of Energy Efficient and Custom AI IC, ShanghaiTech University, Shanghai 201210, China (e-mail: wangcheng1@shanghaitech.edu.cn).}

\thispagestyle{empty}
\pagestyle{empty}

\begin{abstract}
Unmanned aerial vehicles (UAVs) operating in confined, cluttered environments face significant performance degradation due to nonlinear, time-varying unmodeled dynamics—such as ground/ceiling effects and wake recirculation—that are unaccounted for in traditional controllers. While learning-based compensators (e.g., MLPs, TCNs, LSTMs) struggle with historical data dependency, vanishing gradients, and prohibitive computational costs, this work pioneers the integration of a deep photonic reservoir computer (PRC) with feedforward control to overcome these limitations. Harnessing semiconductor laser dynamics and optical feedback, our hardware-implemented deep PRC architecture achieves intrinsic temporal memory without explicit historical inputs,  while reducing training time from hours to milliseconds and slashing inference latency to nanoseconds. Reliable high-performance CFD simulations capturing proximity-induced flows demonstrate that deep PRC delivers residual-force prediction accuracy comparable to or exceeding TCN/MLP baselines, while training only a linear readout layer via ridge regression. By injecting these predictions into a nonlinear feedback PID controller via a feedforward channel, the framework significantly enhances closed-loop tracking stability in confined spaces. Essentially, this work establishes the first deep PRC-based lightweight, ultra-fast solution for real-time UAV dynamic compensation, with promising extensibility to unseen scenarios with more complex fluid environments.
\end{abstract}
\section{Introduction}

Unmanned aerial vehicles (UAVs) have demonstrated tremendous agility and efficacy across diverse domains, ranging from classical monitoring-oriented missions such as wide-area aerial mapping~\cite{Widemapping25} and environmental surveillance~\cite{Rishien24} to more recent and challenging scenarios such as last-mile urban delivery in urban canyons~\cite{Andrew22} and indoor search-and-rescue~\cite{Tomic12} in collapsed buildings. These new scenarios are typically characterized by narrow, gusty, cluttered, unknown, and rapidly changing environments, which impose stringent demands on flight controllers in terms of agility, stability, and resistance to disturbances. Traditional flight controllers that enable UAVs to track desired trajectories in open space under mild-wind conditions generally rely on simplified dynamic models~\cite{Ryll12}. Such models, however, cannot capture the additional nonlinear and time-varying dynamics induced by interactions between the UAV and its surrounding obstacles (typical phenomena include the ground effect~\cite{cheeseman1955,sanchez-cuevas2017} and the ceiling effect~\cite{Hsiao18}), and thereby undermine the deployment of classical controllers in these cases. This motivates and necessitates dynamic compensation, together with feedforward control, which has been supported by many works~\cite{shi19, shiNeuralfly, NeuroBEM21, binlin24, shiNeuralSwarm2PlanningControl2021, yucom23, shiNeuralSwarm20, Lijin23,mlpbansal16, PITCNsaviolo22}, aiming to achieve agility, stability, and feasibility for UAVs in more complex environments. Despite many efforts devoted to this problem, modeling and compensating for the unmodeled dynamics induced by interactions between the UAV and its environment, so that UAVs can achieve stable flight in complex flow fields, remains an open issue.

To enable effective compensation, unmodeled dynamics are commonly treated as an additive residual-force term in the UAV model and mitigated via the feedforward controller~\cite{shiNeuralfly, binlin24, NeuroBEM21}. Traditionally, such a residual force is modeled by a simple empirical model based solely on the UAV’s current state~\cite{cheeseman1955, sanchez-cuevas2017, Hsiao18}, including altitude, vehicle velocity, and vehicle geometry such as rotor diameter and airframe size. However, this residual force is inherently nonlinear, time-varying, and tightly coupled with the vehicle's state over a period of time. Consequently, empirical models are inadequate for use as a feedforward compensation term and fail to deliver satisfactory near-obstacle maneuvering performance for UAVs. To address these limitations, recent works have utilized the neural network (NN)-based model to capture the strong nonlinearity in the dependence of the residual force on the vehicle states. By inputting the historical data, multilayer perceptrons (MLPs)\cite{mlpbansal16} and temporal convolutional networks (TCNs) \cite{PITCNsaviolo22} both demonstrated certain improvements in the residual force prediction. 
However, in this way, the input dimension grows with the window length of historical data, which significantly increases network complexity and training cost. Moreover, deciding the window horizon is a hassle and often requires trial and error. To reduce reliance on explicit historical data length, a parallel line of work has adopted recurrent neural networks (RNNs)~\cite{zhou2021controltailsittervtoluav} and its variants, such as long short-term memory (LSTM)~\cite{chen21} and gated recurrent units (GRUs)~\cite{YANG2024}, which take only the current state as input. Although only the current state is fed at each timestep, the RNN updates a recurrent hidden state that encodes a compact summary of the entire history, so the output implicitly depends on past states. However, these approaches are susceptible to vanishing and exploding gradients~\cite{pascanu13}. Overall, despite the success of NN-based approaches in many nonlinear modeling problems, for the unmodeled dynamic compensation problem addressed in this work, existing NN-based solutions suffer from either the determination of the input length or gradient issues. Furthermore, in practical deployments, these networks face another issue: large network models impose additional overhead on the UAV’s onboard computer, increasing energy usage and ultimately shortening flight endurance.

\begin{figure*}[t]
    \centering
    \includegraphics[width=0.95\linewidth]{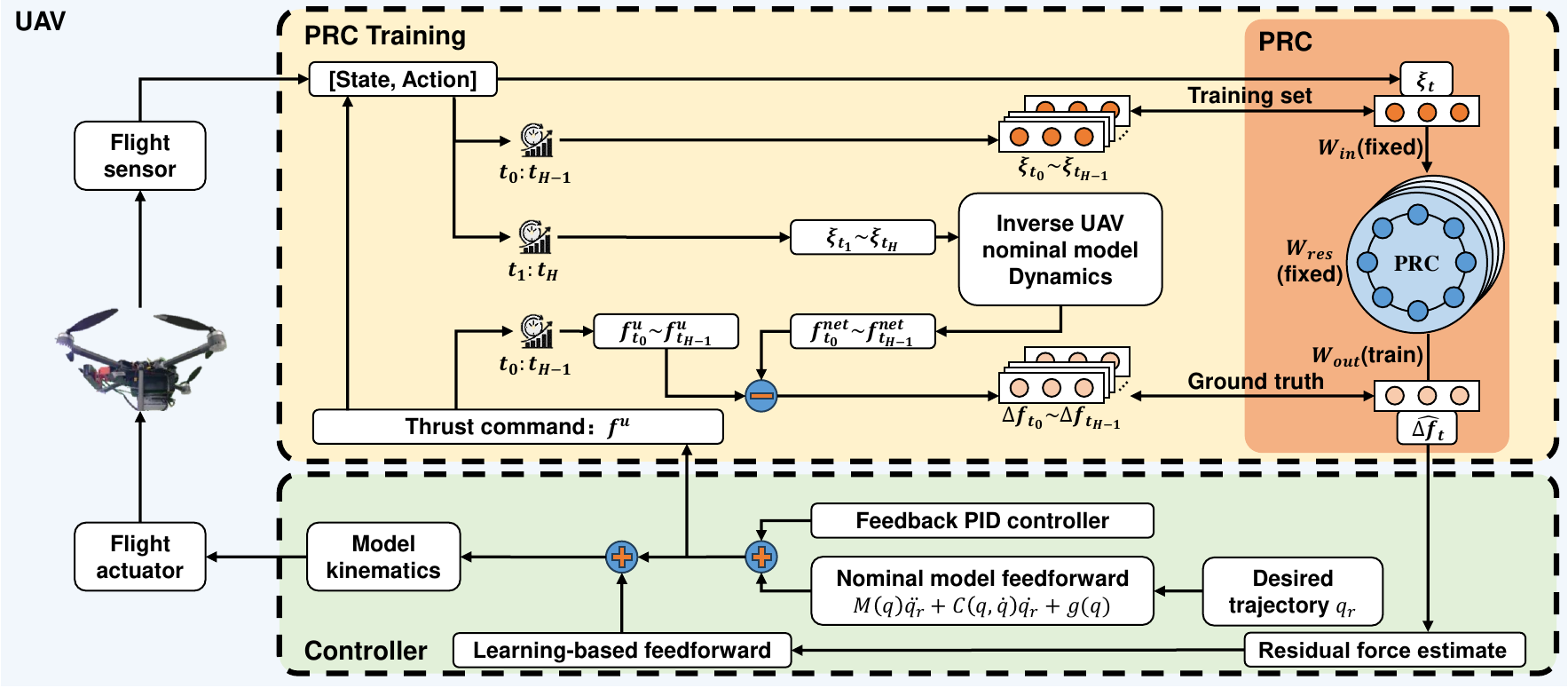}
    \caption{Framework of the deep PRC compensation system. The system includes three parts: training and ground-truth generation; deep PRC inference; Integration of the deep PRC-predicted residual force with the controller.}
    \label{fig:fig1}
\end{figure*}

To address the aforementioned limitations, we propose a novel deep photonic reservoir computer (PRC)-based feedforward controller designed to compensate for nonlinear, time-varying unmodeled dynamics and enable stable flight in complex and confined environments. Deep PRC is a real-time, adaptive form of recurrent neural network (RNN) that has attracted significant attention for its ability to efficiently learn from nonlinear, time-varying sequential data—with applications in chaotic sequence prediction~\cite{Tang22}, nonlinear channel equalization~\cite{Li25}, and spoken digit recognition~\cite{Brunner13}.
The strengths of deep PRC stem from several key characteristics: first, it can be physically implemented in hardware~\cite{Brunner19}, leveraging the nonlinear dynamics of semiconductors rather than relying solely on digital algorithms; second, unlike conventional RNNs, the input and reservoir weights in deep PRC are randomly initialized and fixed, with only the readout layer requiring training; third, since the hidden-layer weights are determined by the physical states of semiconductor lasers, deep PRC naturally avoids the issues of gradient explosion and vanishing that commonly arise in backpropagation-based training. Finally, deep PRC retains the inherent temporal memory of RNNs, making it highly effective for time-series prediction and reconstruction tasks.
Furthermore, compared to von Neumann architecture-based electronic computing, deep PRC provides higher energy efficiency and lower computational latency. Despite these advantages, single-layer PRC has not previously been applied in robotics control, largely due to concerns about its nonlinear fitting capacity. These concerns are addressed in our framework through the use of a deep PRC architecture~\cite{Shen23}.
Extensive evaluations using high-fidelity computational fluid dynamics simulations—assessed in terms of prediction and tracking errors—demonstrate that the deep PRC possesses sufficient learning capability for the targeted dynamic compensation problem. To the best of our knowledge, this represents the first successful integration of deep PRC with a feedforward controller, achieving flight performance comparable to state-of-the-art methods~\cite{sanchez-cuevas2017, PITCNsaviolo22}, while offering orders-of-magnitude faster training and lower inference latency. In our experiments, training a TCN required about 45 minutes and an MLP about 27 minutes, whereas the deep PRC completed training in $\ll 1\,\mathrm{s}$ (milliseconds) by solving only a linear ridge regression; inference per step was approximately $4\,\mathrm{ms}$ for the TCN, $1\,\mathrm{ms}$ for the MLP, and $\ll 1\,m\mathrm{s}$ (microseconds) for the deep PRC. Moreover, owing to its low training complexity and rapid convergence, we demonstrate that our framework can achieve real-time reconfiguration of the readout weights, enabling adaptation to unseen scenarios. In summary, this work not only provides a feasible solution to the challenging unmodeled dynamics compensation problem but also highlights the potential advantages of deep PRC-based solutions, including faster training, reduced inference latency, lower power consumption, and—importantly—strong generalization ability, paving the way for more challenging control tasks in UAVs and other applications.

\section{Preliminary}

RC is a special RNN, developed from echo state networks and liquid state machines. Unlike conventional architectures, RC employs fixed weights in both the input and reservoir layers, where only the readout layer requires training~\cite{Nakajima21}. This design substantially reduces training cost, facilitates online adaptation, and offers strong capabilities for modeling nonlinear dynamical systems. In recent years, physical implementations of RC have reached maturity on several substrates, including memristors, spintronic devices, and certain photonic platforms. PRC leverages the nonlinear dynamics of semiconductor lasers and optical feedback to construct high-dimensional reservoirs in the optical domain. By exploiting ultrafast photonic processes, PRC delivers low latency, high bandwidth, and superior energy efficiency compared with its electronic counterparts, positioning it as a strong candidate for real-time sequence learning and large-scale information processing~\cite{Brunner19}.

In recent years, continual exploration and development of diverse PRC architectures have significantly enhanced their performance and extended their applicability across various domains. Early developments in PRC explored two technical routes: spatial architectures~\cite{Vandoorne14} and time-delay architectures~\cite{Appeltant11}. Owing to its compact footprint and ability to emulate a large number of neurons, time-delay architectures have become the dominant paradigm. Building upon this foundation, Tang et al.~\cite{Tang23, Hulser22} demonstrated the feasibility of asynchronous reservoirs, offering greater flexibility in hardware deployment. Parallel reservoirs were later introduced~\cite{Li23, Tang22}, leveraging wavelength-division multiplexing to enable simultaneous multi-channel input, thereby enhancing scalability for high-throughput tasks. Most notably, in 2023, Shen et al.~\cite{Shen23} proposed the deep reservoir architecture, which significantly enhanced the computational power of PRC, addressing one of the key limitations of photonic computing. With the combined advantages of compactness, deployment flexibility, multi-channel input, and enhanced computational capacity, PRC is now particularly well aligned with the stringent requirements of unmodeled dynamics compensation.

PRC has been extensively investigated in the field of time-series prediction, with studies exploring time-series prediction such as chaotic sequence prediction~\cite{Tang22}. Furthermore, PRC has also been validated in various time-series reconstruction tasks, including optical fiber communication systems for noise mitigation and nonlinear compensation~\cite{Li25}, and speech digit recognition for robust sequence classification~\cite{Brunner13}. These efforts demonstrate PRC’s strong potential for the challenge faced by the learning-based control problem in the robotics community, especially in tasks requiring agile agents. Collectively, these results motivate deep PRC as a promising direction for learning-based dynamic compensation in UAVs, where accurate, low-latency estimation of unmodeled dynamics and feedforward compensation using the predicted residual force are essential in confined, proximity-induced flow regimes. 
\section{Methodology}

Our approach proceeds as follows: We model the additional nonlinear and time-varying dynamics induced by interactions between the UAV and its environment as an additive residual force. Next, a deep PRC-based predictor is employed to map the UAV states and control inputs to this residual and to produce a residual-force estimate. Finally, this estimate is applied as a feedforward term in the controller to improve tracking performance. Figure~\ref{fig:fig1} shows the workflow comprising offline training, online prediction, and feedforward integration.

\subsection{Formulation of Unmodeled Dynamics}

Let the quadrotor state be 
$\mathbf{x} = [\,\mathbf{p}^\top,\mathbf{v}^\top,\mathbf{q}^\top,\boldsymbol{\omega}^\top\,]^\top$, 
where $\mathbf{p}\in\mathbb{R}^3$ is the world-frame position, 
$\mathbf{v}\in\mathbb{R}^3$ the linear velocity, 
$\mathbf{q}\in\mathbb{R}^4$ the unit quaternion describing orientation, 
and $\boldsymbol{\omega}\in\mathbb{R}^3$ the body-frame angular velocity. 
The dynamics are given by
\begin{equation}
\begin{bmatrix}
\dot{\mathbf p}\\[3pt]\dot{\mathbf v}\\[3pt]\dot{\mathbf q}\\[3pt]\dot{\boldsymbol{\omega}}
\end{bmatrix}
=
\begin{bmatrix}
\mathbf v\\[3pt]
\frac{1}{m}\big(\mathbf f_u(\mathbf x)+\Delta\mathbf f(\mathbf x,t)\big)+\mathbf g\\[3pt]
\tfrac{1}{2}\big(\mathbf q \odot \boldsymbol{\omega}\big)\\[3pt]
\mathbf J^{-1}\!\left(\boldsymbol{\tau}_u(\mathbf x)-\boldsymbol{\omega}\times\mathbf J\,\boldsymbol{\omega}\right)
\end{bmatrix},
\label{eq:full_rigidbody}
\end{equation}

where $m>0$ is the vehicle mass, $\mathbf J\in\mathbb{R}^{3\times3}$ the inertia matrix, 
and $\mathbf g=[0,0,-g]^\top$ the gravity vector. 
The operator $\odot$ denotes the quaternion–vector product.  

The total thrust along the body $z$-axis is denoted by $T(\mathbf x)$. 
Rotating this body-frame thrust by the current quaternion yields the world-frame thrust vector,
\vspace{-5pt}
\begin{equation}
\mathbf f_u(\mathbf x) = \mathbf q \odot T(\mathbf x)
= \big[f_{u,x}(\mathbf x),\ f_{u,y}(\mathbf x),\ f_{u,z}(\mathbf x)\big]^\top .
\label{eq:thrust_mapping}
\end{equation}

The term $\Delta\mathbf f(\mathbf x,t)\in\mathbb{R}^3$ is defined as the 
\textit{residual force due to unmodeled dynamics}, which is not captured by the nominal model and varies with time. 
Our objective is to compensate for this residual force through a learned feedforward term to improve tracking accuracy and robustness.

In what follows, we take motion along the vertical axis as an example, since unmodeled dynamics such as ground effect, ceiling effect, and wake interactions can have a significant influence in this direction. These phenomena directly perturb the thrust and the vertical acceleration, making altitude control and closed-loop stability particularly sensitive along this axis. Based on \eqref{eq:full_rigidbody} and \eqref{eq:thrust_mapping}, the discrete-time relations for the vertical motion are:
\begin{align}
z_{t+1} &= z_t + T_s\, v_{z,t},\label{eq:disc3}\\
v_{z,t+1} &= v_{z,t} + T_s\,\dot v_{z,t},\label{eq:disc1}\\
\dot v_{z,t} &= -\,g + \frac{1}{m}\, (f_{u,z,t}(x)+\Delta f_{z,t}(\mathbf{x},t)),\label{eq:disc2}
\end{align}
where $T_s$ denotes the sampling interval. From \eqref{eq:disc1}--\eqref{eq:disc2}, the net force along the $z$-axis at time $t-1$ can be reconstructed using the recorded states at $t$ and $t-1$. Subtracting the commanded thrust at $t-1$ then yields the residual:
\begin{align}
    f^{\mathrm{net}}_{z,t-1} &= m\Big(\tfrac{v_{z,t}-v_{z,t-1}}{T_s}+g\Big),\\
    \Delta f_{z,t-1} &= f^{\mathrm{net}}_{z,t-1}-f_{u,z,t-1}.
\end{align}

This formulation represents the unmodeled dynamics as a residual force and provides training targets for learning. Accurately predicting $\Delta f_{z,t}$, however, remains challenging because the residual force is nonlinear, time-varying, and influenced by complex environments.

\begin{figure}[b]
    \centering
    \includegraphics[width=0.95\linewidth]{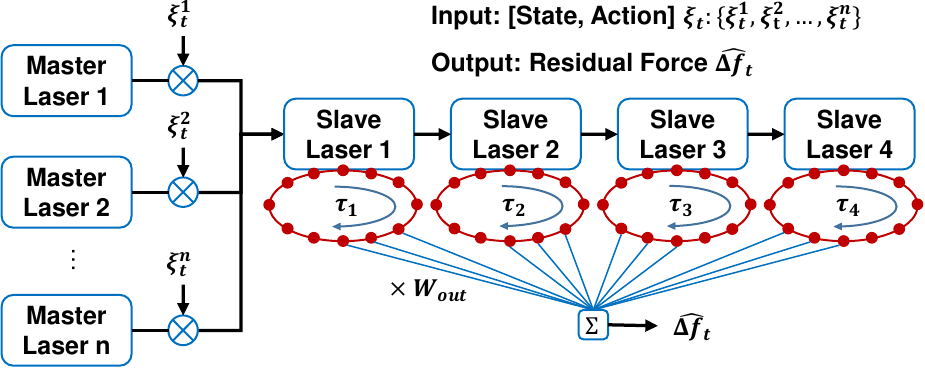}
    \caption{Schematic architecture of the deep PRC. Virtual neurons for the reservoir are generated in each layer through optical feedback delay.}
    \label{fig:fig2}
\end{figure}

\subsection{Deep PRC module: mechanism and interface}

We use a deep PRC module to approximate the residual-force mapping on the vertical axis. The module predicts the vertical residual $\Delta f_{z,t}$ from the input vector.

\[
\xi_t = [\,v_{z,t},\, f_{z,t},\, p_{z,t}\,],
\]
which collects vertical velocity $v_{z,t}$, thrust command $f_{z,t}$, and altitude $p_{z,t}$. This choice is physically motivated: $f_{z,t}$ reflects thrust-dependent effects, $v_{z,t}$ captures flow-related terms linked to induced velocity and recirculation, and $p_{z,t}$ provides altitude information relevant to proximity effects (e.g., ground and ceiling)~\cite{sanchez-cuevas2017,cheeseman1955}.

Figure~\ref{fig:fig2} illustrates the architecture of the deep PRC network. At each time step, the Input vector $\xi_t$ consists of multiple sub-parameters $\xi_1 \sim \xi_n$, which are individually modulated onto $n$ single-mode master lasers. To enable parallel injection, wavelength-division multiplexing (WDM) is employed, with each master laser operating at a distinct wavelength. The modulated optical signals are combined by an optical coupler and unidirectionally injected into the first-layer slave laser (Slave Laser~1) through optical injection locking. A portion of the output from Slave Laser~1 is subsequently injected into Slave Laser~2 in the second layer, also achieving mutual locking. This cascading process continues, with Slave Laser~2 locking with Slave Laser~3, and Slave Laser~3 locking with Slave Laser~4. As a result, the lasing frequencies of all slave lasers are locked to the predefined wavelengths of the master lasers. 

Within each hidden layer, the slave laser is subject to an optical feedback loop, which generates a large number of virtual neurons through nonlinear laser dynamics. The feedback delay in each layer, denoted as $\tau_1 \sim \tau_4$, corresponds to the length of the optical feedback loop. These delays determine the relationship between the neuron time interval $\theta$ and the number of virtual nodes $N$, given by $\tau = \theta \times N$. These neurons interact with one another, thereby performing a high-dimensional mapping of the input data as in Figure~\ref{fig:fig1}:
\[
r_t = PRC\big(W_{in}\xi_t + W_{res}\xi_{t-1}\big).
\]

The mapped features $r_t$ are collected at the readout layer, where the target output is obtained via a weighted summation of all neuron states:  
\[
\widehat{\Delta f}_{z,t+1\mid t} = W_{out} r_t .
\]

The output weights are optimized using ridge regression, ensuring computational efficiency for online training:  
\[
W_{out} = (r_t^\top r_t + \lambda I)^{-1} r_t^\top \Delta f_{z,t+1|t}.
\]

It is worth noting that, since the deep PRC continuously receives the input vector $\xi_t$, the past states inherently affect the dynamical states of the lasers and are thus implicitly stored within the system. Consequently, the deep PRC possesses intrinsic memory, enabling inference without the need to retransmit historical information.

\subsection{Integration with feedforward controller}

For the system (\ref{eq:disc3})–(\ref{eq:disc2}), the control input $f_{u,z}$ is designed as a nonlinear feedback  PID controller~\cite{shiNeuralfly} augmented with a deep PRC-based feedforward compensation, as expressed in (8).

Given the inertia, Coriolis/centrifugal, and gravity terms of the rigid-body dynamics denoted by $M(\mathbf q)$, $C(\mathbf q,\dot{\mathbf q})$, and $g(\mathbf q)$, respectively; the velocity-tracking error $\mathbf{s}=\dot{\mathbf q}-\dot{\mathbf q}_r$; and positive-definite gains $K, K_I$, we define
\begin{equation*}
    u_{NL}(\mathbf x) := M(\mathbf q)\ddot{\mathbf q}_r + C(\mathbf q,\dot{\mathbf q})\dot{\mathbf q}_r + g(\mathbf q) - K\mathbf s - K_I\!\int \mathbf s\,dt ,
\end{equation*}
which is the output of the baseline nonlinear feedback PID controller computed from the current state $(\mathbf q,\dot{\mathbf q})$ and the reference signals $(\mathbf q_r,\dot{\mathbf q}_r,\ddot{\mathbf q}_r)$. The control input applied in (\ref{eq:disc3})–(\ref{eq:disc2}) is then
\begin{equation*}
    f_{u,z,t} = u_{NL}(\mathbf x) - \widehat{\Delta f}_{z,t+1|t}.
\end{equation*}

In the absence of disturbances ($\Delta f=0$), the nominal controller $u_{NL}$ ensures asymptotic tracking for the continuous-time model \eqref{eq:full_rigidbody}–\eqref{eq:thrust_mapping}. For further stability proofs, please refer to~\cite{shiNeuralfly}. Substituting the above $f_{u,z,t}$ into \eqref{eq:disc3}–\eqref{eq:disc2} gives
\begin{align*}
    e_{\Delta,t} :&= \Delta f_{z,t} - \widehat{\Delta f}_{z,t+1|t}, \\
v_{z,k+1} &= v_{z,k} + T_s\!\left[-g+\tfrac{1}{m}\big(u_{NL}(\mathbf x)+e_{\Delta,t}\big)\right].
\end{align*}
If $e_{\Delta,t}$ is bounded, the closed loop is input-to-state stable (ISS) with respect to $e_{\Delta,t}/m$, and the steady-state tracking error scales with the prediction error. If $e_{\Delta,t}\to 0$, the residual vanishes, the dynamics reduce to the disturbance-free case, and the tracking error converges to zero.

\begin{figure*}[t]
    \centering
    \includegraphics[width=0.95\linewidth]{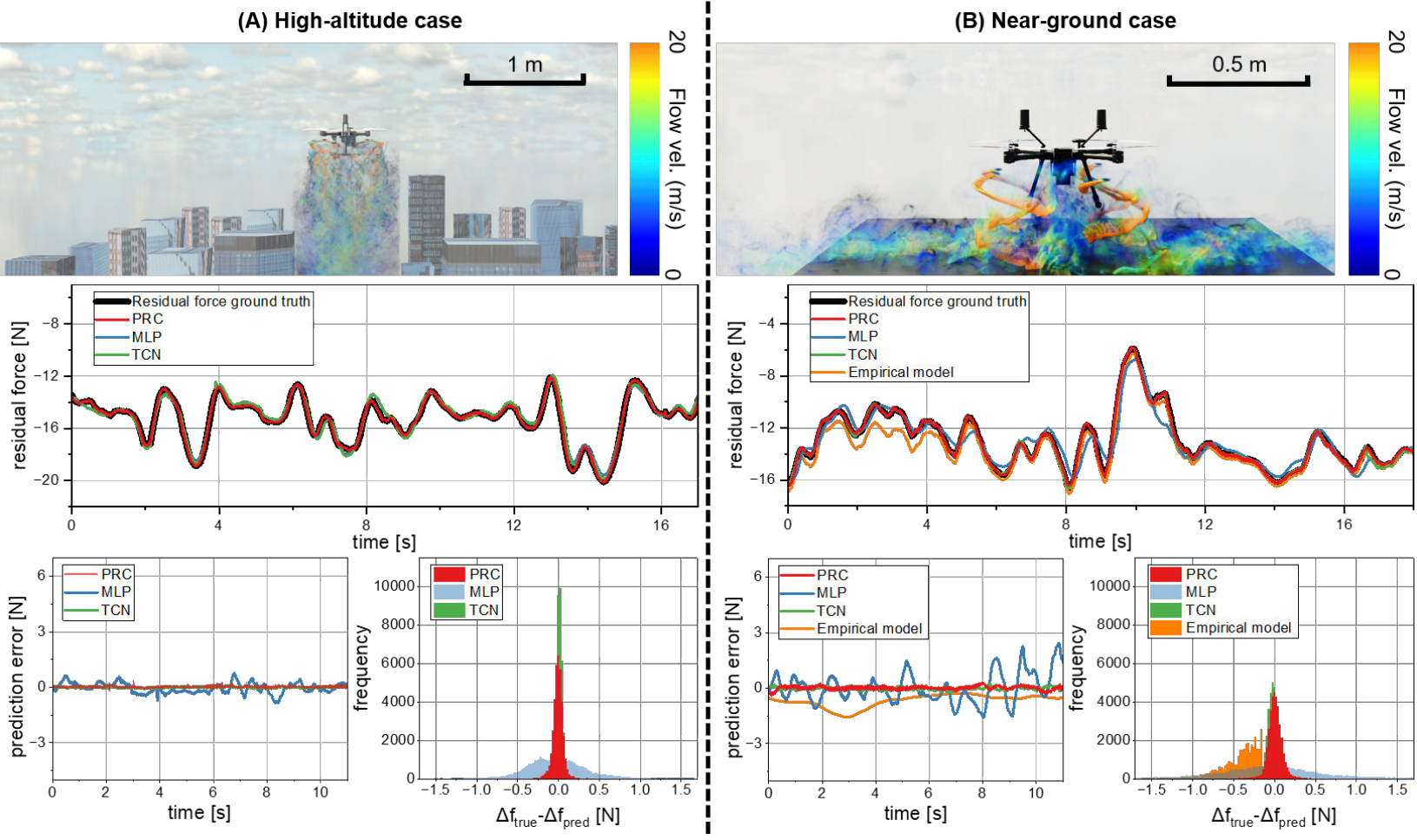}
    \caption{Comparison of residual force prediction in two flight cases:\textbf{(A) high-altitude case} and \textbf{(B) near-ground case}. \textbf{Top}: CFD flow visualizations around UAV, with stronger recirculation and vortex structures near the ground. \textbf{Middle}: ground truth residual force versus prediction from deep PRC and baselines. \textbf{Bottom}: prediction error and error distribution of $\Delta f_{true}-\Delta f_{pre}$. Across both cases, deep PRC matches TCN in delivering the best prediction performance.}
    \vspace{-5pt}
    \label{fig:fig3}
\end{figure*}
\section{CFD Simulation System for UAV}

To evaluate the proposed compensator under realistic aerodynamic conditions, we employ a CFD-based fluid–structure simulator~\cite{UAVHybridSim2025}\footnote{Please refer to \url{https://collisionmodel.com} for more details and updates (accessed Mar. 6, 2026).}. The airflow is assumed subsonic and modeled by the unsteady isothermal weakly compressible Navier–Stokes equations~\cite{anderson2010aero}:
\begin{align}
\frac{\partial \rho}{\partial t} + \nabla \cdot (\rho \mathbf{u}) &= 0, \\[4pt]
\frac{\partial (\rho \mathbf{u})}{\partial t} + \nabla \cdot (\rho \mathbf{u}\mathbf{u}) &= -\nabla p + \nabla \cdot \boldsymbol{\sigma} + \mathbf{F}, \\[4pt]
p &= \rho R T_0, \\[4pt]
\boldsymbol{\sigma} &= 2\rho \nu \mathbf{S} - \rho \nu' (\nabla \cdot \mathbf{u}) \mathbf{I}.
\end{align}

Here, $\rho$ is density, $\mathbf{u}$ velocity, $p$ pressure, $\mathbf{F}$ the external force field, $R$ the gas constant, $T_0$ the reference temperature, $\nu$ and $\nu'$ the kinematic and bulk viscosities, $\mathbf{S}$ the strain-rate tensor, and $\mathbf{I}$ the identity matrix.

We solve these airflow dynamics using the lattice Boltzmann method (LBM), leveraging its conservative local updates and low numerical dissipation. These properties yield high efficiency when optimized on GPUs and support large domains and multi-UAV scenarios. In our setting, the solver captures key aerodynamic phenomena relevant to confined and cluttered environments, including ground and ceiling effects, wake recirculation, and obstacle-induced proximity flows. This configuration provides sufficient fidelity for control evaluation and is efficient for closed-loop testing of learning-based compensators.

\section{Experiment Setup}
\subsection{System Setup}
The CFD fluid–structure UAV simulator was executed on a workstation with an NVIDIA GeForce RTX 4070 with 12\,GB VRAM. The flow solver was configured with a spatial resolution of 0.04\,m and a maximum flow speed of 40\,m/s. The simulated vehicle was an Amovlab P600 quadrotor with mass 3.3\,kg, wheelbase 600\,mm, and rotor radius 0.19\,m. The onboard stack Amovlab Allspark NX with PX4 was emulated with a 100\,Hz control loop. The simulator provides higher-accuracy state estimates than the real system, offering a precise basis for evaluating dynamics compensation.

\subsection{Data Collection}


We evaluate the deep PRC for dynamics compensation in two flight regimes: high-altitude free flight and low-altitude flight with ground effect. Each test lasts $200$\,s with speed and acceleration limits $v_{\max}=3.4$\,m/s and $a_{\max}=10.7$\,m/s$^{2}$. Each trajectory is flown for ten independent trials. For closed-loop validation, we use a confined environment with a ground plane and a pipe/duct structure that induces strong proximity-induced flow interactions, stressing the controller under adverse conditions.

\begin{enumerate}
    \item \emph{Multisin trajectory.} 
    $z_d(t)=\sum_{i=1}^{N} A_i \sin\!\bigl(2\pi f_i t + \phi_i\bigr)$, with frequencies $[0.04,\,0.10,\,0.13,\,0.17,\,0.37]$\,Hz, amplitudes $[2.0,\,1.5,\,0.97,\,0.5,\,0.33]$\,m, and phases $[0,\,\pi/4,\,\pi/2,\,\pi/3,\,\pi/6]$. 
    In the ground-effect regime, the same $f_i$ and $\phi_i$ are used with reduced amplitudes $[0.4,\,0.3,\,0.19,\,0.1,\,0.06]$\,m to match the smaller safe altitude range.
    
    \item \emph{Random trajectory.}
    The altitude bounds are $0$--$10$\,m. At each control step, the next desired altitude is sampled within $\pm 1.5$\,m of the current value. If a bound is reached, the direction is reflected to keep $z_d(t)$ within limits.
\end{enumerate}

Across all experiments, two-thirds of the data are used for training and one-third for validation.

\subsection{Deep PRC Setup}
In the simulation experiments, deep PRC adopted a 3-channel parallel architecture with 4 deep reservoir layers, each containing 50 nodes. The injection ratio was $-5$\,dB, and the feedback ratio was $-30$\,dB. In deep PRC, the laser-related parameters play a role analogous to hyperparameters in conventional neural networks. It is important to note that these operational parameters were not quantitatively optimized in this study.

\subsection{Baselines}

After configuring deep PRC, we compared it against two learning-based predictors and one empirical model. 
The first baseline is a memoryless multilayer perceptron (MLP) with no historical inputs. The second is a temporal convolutional network (TCN) that ingests a fixed-length history of $T=10$ samples at $\Delta t=0.009$\,s. 
The TCN has four 1D convolutional layers (16 channels each) followed by a three-layer MLP of sizes $64$–$32$–$32$, whereas the MLP baseline has four hidden layers of sizes $64$–$32$–$32$–$32$. 
All layers in both networks use ReLU activations, batch normalization, and $10\%$ dropout. 
Both networks are trained on collected data for $200$ epochs with Adam (batch size $64$, constant learning rate $10^{-3}$). 

As a non-learning baseline, we include the classical ground-effect model~\cite{sanchez-cuevas2017}, used as a thrust multiplier $\eta_{\mathrm{GE}}(z;R,d)=T_{\mathrm{IGE}}/T_{\mathrm{OGE}}$. 
In implementation, we form a residual $\Delta F_z=(\eta_{\mathrm{GE}}-1)F_{\mathrm{des}}$, with $\eta_{\mathrm{GE}}\!\to\!1$ at large $z$. This analytical model is inapplicable at high altitudes where ground effect vanishes; accordingly, we exclude it from the high-altitude comparisons. Together, these baselines contrast a memoryless predictor (MLP), a sequence-based predictor using fixed history (TCN), and a near-ground analytical prior, enabling a balanced assessment of deep PRC in both prediction and closed-loop settings.

\section{Results and Discussion}

We validate the effectiveness of our approach from three aspects. First, we evaluate prediction accuracy by comparing deep PRC against baselines across high-altitude and near-ground regimes. Second, we demonstrate closed-loop capability by injecting deep PRC’s predicted compensation as a feedforward term and measuring tracking performance and stability in a confined scene with a ground plane and a pipe. Finally, we present the additional advantages of deep PRC by analyzing runtime, latency, and training cost, and discuss the feasibility of online training and onboard adaptation.

\begin{figure}[t]
    \centering
    \includegraphics[width=1.0\linewidth]{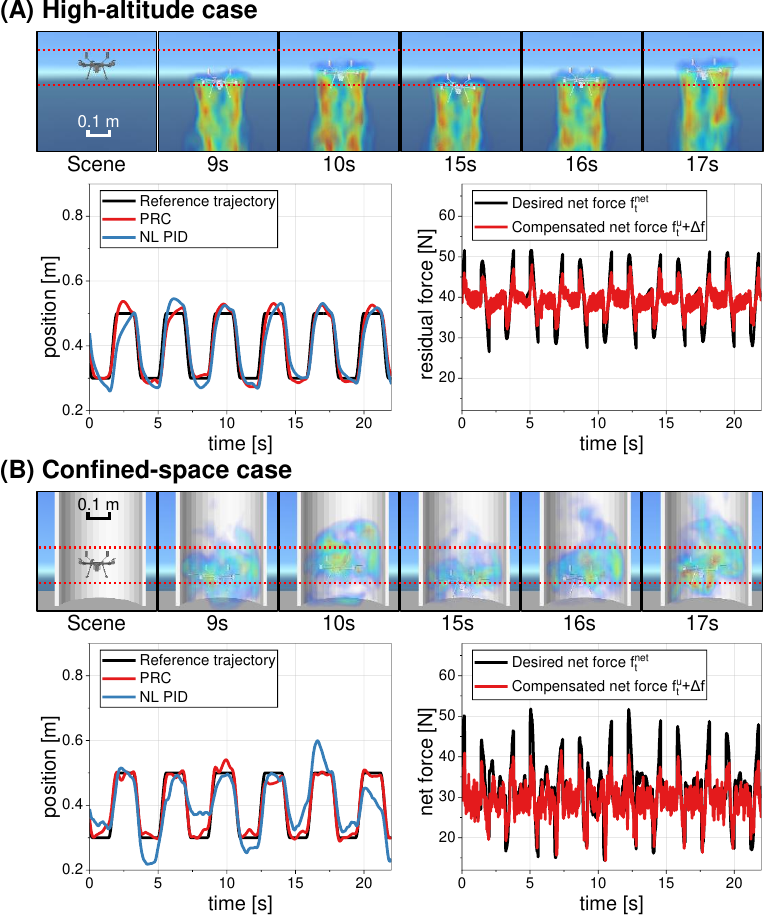}
    \caption{Closed-loop trajectory tracking in two cases: \textbf{(A)} high altitude and \textbf{(B)} confined space. For each case, \textbf{Top:} CFD flow snapshots along the closed-loop trajectory of a UAV under PRC-based feedforward control. The vehicle’s center of mass is bounded by red dashed lines denoting the desired trajectory limits. \textbf{Bottom:} comparison of \textbf{(left)} position under deep PRC feedforward and nonlinear feedback PID controllers against the reference trajectory, and \textbf{(right)} net force compensated from deep PRC compared with the desired net force.}
    \label{fig:fig4}
\end{figure}

\subsection{Prediction Performance}
\label{paper:6A}
Figure~\ref{fig:fig3} compares predicted residual force with the ground truth in two representative cases: high altitude flight on the left and flight near the ground with ground effect on the right. We select ground effect because it is a canonical and safety-critical proximity regime in low altitude operations, and it has a widely used conventional model that serves as a clear reference. As the conventional baseline, we use the ground-effect model~\cite{sanchez-cuevas2017}, which is included only in the near-ground case, as specified in the experiment setup.

For both cases in the middle of Fig.~\ref{fig:fig3}, deep PRC predicts the residual force caused by external dynamics more accurately than the MLP and the conventional model. Its accuracy is comparable to that of the TCN trained with a series of historical data, while deep PRC does not require explicit historical inputs. In the high altitude case, all methods capture the overall trend, and deep PRC yields smaller instantaneous errors. In the ground effect case, prediction becomes more difficult because nonlinear and unsteady aerodynamics are stronger, and past states influence the dynamics. The conventional model cannot capture these effects, and the MLP degrades, whereas deep PRC maintains accuracy comparable to the TCN without using history inputs.

As shown at the bottom of Fig.~\ref{fig:fig3}, we also present prediction error and error distributions for the same trials. Deep PRC errors cluster tightly around zero with low variance in both regimes, and the TCN shows a similarly tight distribution. The MLP and the conventional baseline display wider distributions with heavier tails. Although errors increase for all methods near the ground, deep PRC remains narrower than the baselines, which indicates stronger robustness across operating conditions. These results show that deep PRC reaches performance on par with TCN without long historical inputs, and it yields higher accuracy than memoryless and conventional baselines.

\begin{figure}[t]
    \centering
    \includegraphics[width=0.8\linewidth]{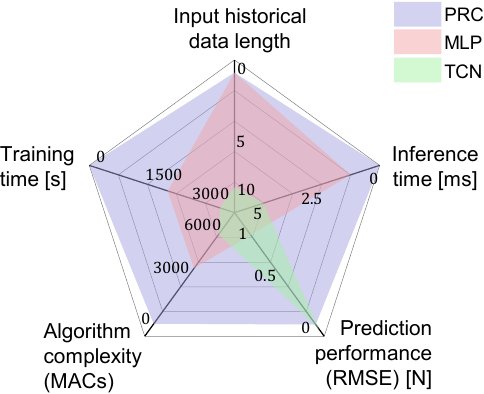}
    \caption{Radar-plot comparison of deep PRC, MLP, and TCN across training time, inference time, MACs, input historical length, and prediction accuracy. Deep PRC achieves the lowest complexity and latency while maintaining accuracy comparable to TCN.}
    \label{fig:fig5}
\end{figure}


\subsection{Closed-Loop Tracking Performance}
Figure~\ref{fig:fig4} reports closed-loop tracking performance when the deep PRC-predicted residual force is injected as a feedforward term. In high-altitude flight, the disturbance is weak, the flow field distribution is uniform, and the compensated and uncompensated trajectories are almost identical. A post-run check confirms that the commanded compensation matches the residual required by the dynamics. In the proximity case, environment-induced external dynamics become significant. Flow field snapshots around the UAV at different times show that, even at the same position and altitude, the local flow distribution differs markedly from one instant to the next, which leads to different disturbances. Without feedforward, the tracking performance degrades. With deep PRC feedforward, the trajectory progressively aligns with the reference, and the commanded compensation matches the residual required by the dynamics. These results indicate that the compensator improves closed-loop performance in complex environments.

\begin{figure}[b]
    \centering
    \includegraphics[width=0.95\linewidth]{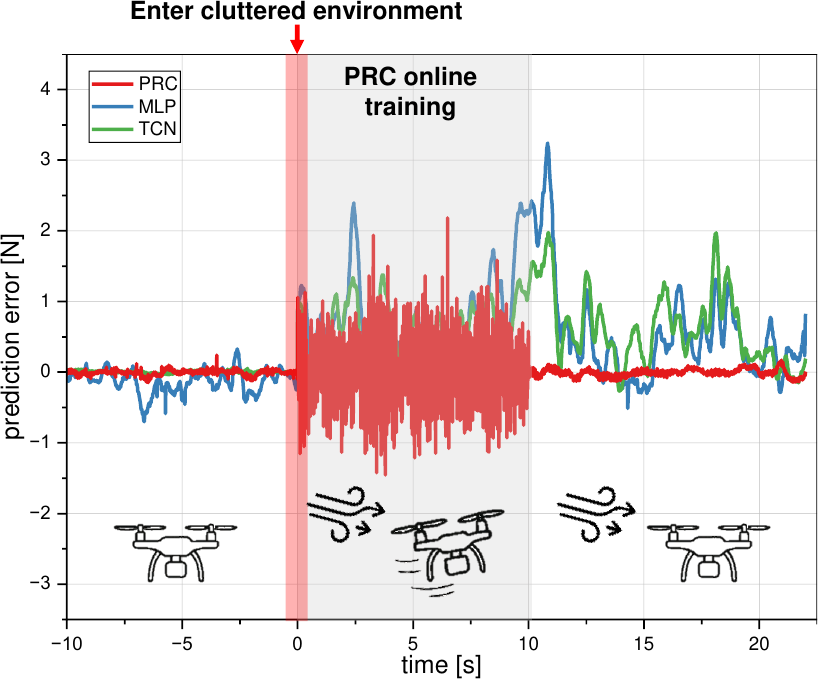}
    \caption{Prediction error of deep PRC, MLP, and TCN during flight. At $t=0$ s, the UAV enters a cluttered environment. Without online adaptation, MLP and TCN degrade, while deep PRC adapts after 10 s of online training and stabilizes at low error.}
    \label{fig:fig6}
\end{figure}

\subsection{Discussion}

\subsubsection{Multi-dimensional Comparison}
As established in~\ref{paper:6A} of this section, deep PRC achieves accuracy comparable to a TCN trained with a series of historical data while outperforming MLP and the conventional ground-effect model. Figure~\ref{fig:fig5} extends the comparison beyond accuracy to four dimensions: inference time, training time, required history length, and algorithmic complexity (evaluated by multiply accumulate operations, MACs). With approximately $10^{5}$ samples, the TCN trains in about 45 minutes and the MLP in about 27 minutes, whereas deep PRC completes training within seconds because the reservoir extracts features at hardware speed and training reduces to a single ridge-regression solve. As for inference time, MLP and TCN operate at the millisecond scale per step, while deep PRC runs at the nanosecond scale owing to its computation directly in the optical domain. Deep PRC does not require explicit historical data as input, as its intrinsic reservoir dynamics inherently provide memory for temporal processing, while still achieving accuracy comparable to that of TCNs trained on extended historical sequences. The algorithmic complexity also favors deep PRC, since only the linear readout layer is trained and executed digitally.


\subsubsection{Scenario for Online Training}
Figure~\ref{fig:fig6} shows why ultra-fast training and inference are critical for onboard adaptation. Before $t=0~s$, the vehicle flies in a known condition, and all predictors are accurate. At $t=0~s$, it enters a cluttered environment; prediction errors rise for every method. Over the next $10~ s$, the vehicle collects data at 70~Hz (about 700 samples). Deep PRC then updates its readout with a fast ridge-regression step ($<1\,ms$) while maintaining $\mu s$-scale inference, after which its error returns to a low level. MLP and TCN cannot be retrained within this short window and remain inaccurate. These results highlight that deep PRC enables practical online reconfiguration during flight when conditions change abruptly.

\subsubsection{Transferability}
Beyond UAVs, the proposed dynamics-compensation framework can be extended to other robotic platforms. Robotic systems are commonly modeled using Euler–Lagrange rigid-body dynamics, while their behavior is further influenced by external factors such as friction, contact, wind, and flow. The deep PRC module can be deployed as a feedforward compensator across legged robots, manipulators, other aerial vehicles, and underwater platforms to counter environment-induced residual forces and torques.
\section{Conclusion}

We addressed unmodeled external dynamics in confined, complex environments by integrating a deep photonic reservoir computer with a feedforward compensator for UAV control—the first such integration to our knowledge. In a high-fidelity CFD simulation, deep PRC generated residual-force estimates that, when injected via the feedforward channel, improved closed-loop tracking and stability. Its prediction accuracy matched or exceeded MLP and TCN baselines, while requiring training only of a linear readout. The approach is computationally lightweight: training completes in $<1 ms$ and per-step inference in $<1 \mu s$, thereby reducing computational load and power consumption. Future work includes implementing hardware-in-the-loop online training to realize closed-loop adaptation and transferring the approach to real vehicles through integration on a UAV platform and evaluation in flight.

\section{Acknowledgement}
The authors thank the FlareLab at the School of Information Science and Technology, ShanghaiTech University, for providing the CFD-based fluid-structure simulator and for their valuable advice and technical support throughout this work.
{
    \small
    \bibliographystyle{IEEEtran}
    \bibliography{main}
}

\end{document}